\documentclass[12pt]{article}
\usepackage{amsfonts}
\newcommand{\half}{\frac{1}{2}}

\newcommand{\nn}{\nonumber}

\def\ZZ{{\mathbb Z}}

\begin{document}
\pagestyle{empty}

\begin{center} 
  {\Large \textsf{
      Integrable Ladder t-J Model with Staggered Shift of                    
      \\[4mm]
      the Spectral Parameter}} 
  
\vspace{36pt}
{\bf J.~Ambjorn$^{a,}$\footnote{e-mail:{\sl ambjorn@alf.nbi.dk}}},
{\bf D.~Arnaudon$^{b,}$\footnote{e-mail:{\sl arnaudon@lapp.in2p3.fr}}},
{\bf A.~Sedrakyan$^{a,b,}$\footnote{e-mail:{\sl sedrak@lx2.yerphi.am,
Permanent address: Yerevan Physics Institute, Armenia}}},
{\bf T.~Sedrakyan$^{a,b,}$\footnote{e-mail:{\sl tigrans@moon.yerphi.am,
Permanent address: Yerevan Physics Institute, Armenia}}},
{\bf P.~Sorba$^{b,}$\footnote{e-mail:{\sl sorba@lapp.in2p3.fr}}}\\

\vfill

\emph{$^a$Niels Bohr Institute, Blegdamsvej 17, Copenhagen, 2100 Denmark}
\\[4mm]
\emph{$^b$Laboratoire d'Annecy-le-Vieux de Physique Th{\'e}orique LAPTH}
\\
\emph{CNRS, UMR 5108, associ{\'e}e {\`a} l'Universit{\'e} de Savoie}
\\
\emph{BP 110, F-74941 Annecy-le-Vieux Cedex, France}
\\

\vfill
{\bf Abstract}

\end{center}

The generalization of the Yang-Baxter equations (YBE) in the presence
of $\ZZ_2$ grading along both chain and time directions is presented
and an integrable model of $t-J$ type 
with staggered disposition along a chain of shifts of the spectral 
parameter is constructed. The Hamiltonian of the model is computed in 
fermionic formulation. It involves three neighbour site interactions
and therefore can be considered as a zig-zag ladder model. The Algebraic
Bethe Ansatz technique is applied and the eigenstates, along with
eigenvalues of the transfer matrix of the model are found.
In the thermodynamic limit, the lowest energy of the
model is formed by the quarter filling of the states by fermions
instead of usual half filling.

\vfill
\rightline{LAPTH-804/00}
\rightline{hep-th/0006243}
\rightline{June 2000}

\newpage
\pagestyle{plain}
\setcounter{page}{1}

\section{Introduction}
\indent

The interest to ladder type models was raised in a beginning of
90-s (see for a review \cite{Rice}) in connection with high
temperature superconductivity problems in metal oxides. It is
believed that quasi-one dimensional multi-ladder chains of
strongly interacting electrons reflects the most important
aspects of two dimensional systems and also can reveal some 
properties of the week coupling between conducting planes.

Recently there has been considerable interest in the construction
of integrable ladder type models motivated by the desire to use
powerful technique of Algebraic Bethe Ansatz $(ABA)$ \cite{Bax,FT} 
in the exact investigations of the variety of physical phases
of the models.

In the articles \cite{Wa} integrable ladder models were
constructed by extension of symmetry algebra, in \cite{Ko}
by defining first the ground state and then formulating a model,
which has it. The higher conservation laws of an integrable models,
which contains next to nearest neighbour interactions, were
used in construction of ladder models in \cite{Muu} by developing
the approach of the article \cite{Maj}. There are also some other 
type attempts in this area \cite{Al,For}.

Usually integrable models are homogeneous along a chain, namely,
the spectral $u$ and model parameters are the same in the
product of $R$-matrices along a chain. It is obvious, that if one 
considers arbitrary shifts of the spectral parameters by
some $z_i$ in the monodromy matrix we still will have an integrable
model. But in order to have a local Hamiltonian we need to consider
shifts with the fixed periodicity $n$, which will cause the
interaction of spins (or electrons) within an amount of $n$ neighbours
leading to $n$-ladder model. This type of model was first
considered in \cite{Destri}, developing the ideas
of the articles \cite{Resh} (see also \cite{FLec}), then in a
chain of articles \cite{ZV,ZV1, FR}.

In the article \cite{APSS} we have proposed an inhomogeneous model
based on $XXZ$ spin-chain, where inhomogeneouty appeared not
only in the staggered shifts of the spectral parameter, but
also in change of structure of $R$-matrices in a product along
a chain. Namely, two monodromy matrices of chains $M_s, s=0,1$ were
considered along a time direction, where the $R$-matrices in the 
product have an alternating disposition of the anisotropy
parameter $\pm \Delta$ of the $XXZ$ model.
Besides that, and contrary to case considered in \cite{FLec,ZV,ZV1,FR},
the spectral parameter of the second line has an opposite sign.
Due to the double space translational invariance the Hamiltonian of 
the model contains interaction between three neighbours sites
of the chain and  therefore represents a zig-zag type ladder
model. At the free fermionic point $\Delta = 0$ the model becomes
 a model of two noninteracting fermions, hopping separately
in the odd and even sites of the chain.

In this article we are extending our approach to $spin-1$ case
and considering the t-J model \cite{AK,U,L,S,KOR,BBO,EK}. 
Following \cite{APSS}
we are writing two Yang-Baxter equations $(YBE)$ (see equations \ref{R3}
and \ref{R4} below) for each step of two alternating $R$-matrices
in the chain, but as it appeared in the solution, in this isotropic
in the spin space model two intertwining $R$-matrices are coinciding.
As a result we are being left with the alternating shift of the 
spectral parameter and the change of sign of the spectral parameter in
 second line. As we will see, though this gives us the same Bethe 
Equations $(BE)$ for the spectral parameters of the excitations
as for the model derived in \cite{ZV,ZV1,FR}, 
their energy and the energy of ground state are different. It appears
that in the thermodynamic limit the lowest energy of the model is
reached by the quarter filling of the states by  
fermions, instead of the usual half filling. 

In Section 3 we formulate the model and find a local Hamiltonian,
which has a zig-zag ladder form. It consists of two chains with $t-J$
type Hamiltonians on each of them, the hopping term of electrons
from one chain to the other and two type of interaction terms between 
chains. The first interaction term has the form of spin-spin 
interaction, where one spin is composed from two fermions 
on the same site of the chain, while the other spin is composed
from two different fermions on the neighbour sites of the other chain
of the ladder. The second interaction term has a topological form
of interacting spins and is written for the triangles consisting
of the zig-zag rungs.

In Section 4 we apply $ABA$ in order to find the eigenvalues and 
the eigenstates of the model. At the end of Section we find the ground
state energy of the model and the spectrum of excitations in 
thermodynamic limit.


\section{The Yang-Baxter Equations and its solution.}
\indent

Following \cite{APSS} let us define the monodromy operator of the model as a
product of the two chain monodromy operators $M_0(u)$ and $M_1(u)$
\begin{equation}
\label{M1}
M(u)=M_0(u)M_1(u),
\end{equation}
where 
\begin{eqnarray}
\label{M2}
M_1(u)&=&\bar{R}_{01}^{\iota_2}(u) {R_{02}}^{\iota_1\iota_2}(u) \bar{R}_{03}^{\iota_2}(u)
....\nonumber\\ 
M_0(u)&=&R_{01}(u) \bar{R}_{02}^{\iota_1}(u)R_{03}(u)....
\end{eqnarray}
and we have two operations $\iota_1$ and $\iota_2$,  defined with property 
${\iota_1}^2={\iota_2}^2=1$.

The transfer matrix of the model will be defined as a trace over auxiliary 
spaces $0$ and $0'$ 
\begin{equation} 
\label{T1}
\tau(u)=Tr_0Tr_{0'}M(u)=\tau_0(u)\tau_1(u).
\end{equation}

As it was shown in the article \cite{APSS} the commutativity of 
transfer matrices $\tau(u),\tau(v)$ for arbitrary values of the
spectral parameters u and v can be ensured by the following set of 
Yang-Baxter Equations
\begin{equation}
\label{R3}
\check{R}_{12}(u,v)\check{\bar{R}}_{23}^{\iota_2}(u)\check{R}_{12}(v)=
\check{R}_{23}^{\iota_2}(v)\check{\bar{R}}_{12}(u)\check{\tilde{R}}_{23}(u,v),
\end{equation}
and
\begin{equation}
\label{R4}
\check{\tilde{R}}_{12}(u,v)\check{R}_{23}^{\iota_1\iota_2}(u)
\check{\bar{R}}_{12}^{\iota_1}(v)
=\check{\bar{R}}_{23}^{\iota_1\iota_2}(v)\check{R}_{12}^{\iota_1}(u)
\check{R}_{23}(u,v),
\end{equation}
where we have used the braid (check) formalism for the convenience.

It is also convenient to use the fermionic operator formalism for 
R-matrices and the YBE developed in \cite{HS, AK, GM, US}.

For the $t-J$ model under consideration \cite{AK,U,L,S,KOR,BBO,EK} we 
have to
realize the spin 1 algebra in the fermionic approach. A minimum of two
sorts of fermions is needed in order to express three basic states 
$\mid+\rangle,\mid 0 \rangle,\mid -\rangle$ of the $spin-1$
particle with the $z$ component of the spin equal to $1$, $0$, $-1$ 
correspondingly.

Now let's define ${c_\sigma}^+$, $c_\sigma$, where $\sigma=\uparrow
\downarrow$, as a creation-annihilation operators of fermions with the
 up and down spins respectively, together with their Fock space 
$\mid0\rangle$,$ \mid{\sigma}\rangle$.

The states with definite third projection of the algebra $SU(2)$
can be realized through fermionic Fock space as follows
\begin{equation}
\label{S3}
\mid-\rangle \equiv \mid0 ,\downarrow \rangle, 
\mid+\rangle \equiv \mid \uparrow ,0 \rangle,
\mid0\rangle \equiv \mid0 ,0\rangle,
\end{equation}
numerated as $\mid 1 \rangle,\mid 2 \rangle,\mid 3 \rangle$ respectively.

As it is obvious from the formulas \ref{S3}, we have constructed a
graded space with the following parities for the basic vectors
\begin{equation}
\label{S4}
p(\mid+\rangle)=p(\mid -\rangle)=1, \qquad p(\mid 0\rangle)=0.
\end{equation}

In order to proceed further and write the fermionic $R$-matrix we should 
calculate the Hubbard operator
${X}_m^n=\mid m\rangle\langle n\mid ; \qquad m,n=1,2,3$ 
\begin{eqnarray}
\label{X1}
X_{ m}^k &=&\left( 
\begin{array}{lll}
|-\rangle \langle -|&\qquad |- \rangle \langle +| 
&\qquad |-\rangle \langle 0| \\
|+\rangle \langle -|&\qquad |+\rangle \langle +| 
&\qquad |+\rangle \langle 0|\\ 
|0\rangle \langle -|&\qquad |0\rangle \langle +| 
&\qquad |0\rangle \langle 0|  
\end{array}
\right) \nonumber\\
&=&\left( 
\begin{array}{lll}
(1-n_{\uparrow}) n_{\downarrow}&c^{+}_{\downarrow} c_{\uparrow}
 &(1-n_{\uparrow})c^{+}_{\downarrow} \\
c^{+}_{\uparrow} c_{\downarrow}& n_{\uparrow}(1- n_{\downarrow})
& c^{+}_{\uparrow}(1-n_{\downarrow}) \\ 
(1-n_{\uparrow})c_{\downarrow} & c_{\uparrow}(1-n_{\downarrow})
&(1-n_{\uparrow})(1- n_{\downarrow}) 
\end{array}
\right),
\end{eqnarray}

The trace of this operator is
\begin{equation}
\label{D}
\Delta=X_m^m=1-n_{\uparrow}n_{\downarrow},
\end{equation}
which is an identity operator on the space of states,where the
double occupancy of the sites by fermions is excluded.

Following \cite{AK} let's write down the fermionic $R$-operator
for the t-J model ($spin$ $1$ \cite{U,L,S} model).
\begin{equation}
\label{R5}
\check{R}_{i,j}(u)=a(u)I_{i,j}+b(u)\Pi_{i,j}=a(u)I_{i,j}+b(u)\sum_{m,n=1}^N
(-1)^{p(m)}{X_i}_{n}^m{X_j}_m^n,
\end{equation}
where ${\Pi}_{i,j}$ is the graded permutation operator of the spaces $V_i$ and
$V_J$.

Now by putting the $R$-matrix form (\ref{R5}) into the YBE (\ref{R3})
and (\ref{R4}) and after some calculations one can find $12$ equations
which require the operations $\tilde{}$ and transformation $\iota_2$ to be
\begin{eqnarray}
\label{AB1}
\tilde{a}(u,v)=a(u,v),& &  \tilde{b}(u,v)=b(u,v), \nonumber\\
\frac{a^{\iota_2}(u)}{b^{\iota_2}(u)}&=&\frac{a(u)}{b(u)}.
\end{eqnarray}

The conditions (\ref{AB1}) are reducing the $12$ equations to
following two equations
\begin{eqnarray}
\label{AB2}
a(u,v) [\bar{a}(u)b(v)-\bar{b}(u)a(v)]+b(u,v)\bar{a}(u)a(v)&=&0 
\nonumber\\
a(u,v) [a^{\iota_1}(u)\bar{b}^{\iota_1}(v)-b^{\iota_1}(u)\bar{a}^{\iota_1}(v)]+
b(u,v)a^{\iota_1}(u)\bar{a}^{\iota_1}(v)&=&0
\end{eqnarray}
the consistency condition of which can be found easily as follows
\begin{eqnarray}
\label{AB3}
\frac{b(v)}{a(v)}-\frac{\bar{b}^{\iota_1}(v)}{\bar{a}^{\iota_1}(v)}=
\frac{\bar{b}(u)}{\bar{a}(u)}-\frac{b^{\iota_1}(u)}{a^{\iota_1}(u)}=
constant=\theta.
\end{eqnarray}
Here $\theta$ is the new parameter of our model.

Then the solution of (\ref{AB2}) for the intertwinner parameters 
$a(u,v)$ and $b(u,v)$ will be
\begin{equation}
\label{AB4}
\frac{b(u,v)}{a(u,v)}=\frac{\bar{b}(u)}{\bar{a}(u)}-\frac{b(v)}{a(v)}
\end{equation}

We should now define the $\iota_1$ operation. It is easy to see from
the expression (\ref{AB3}) for $v=u$ that $\iota_1$ operation can be consistently 
defined as follows
\begin{equation}
\label{AB5}
\frac{b^{\iota_1}(u)}{a^{\iota_1}(u)}=-\frac{b(u)}{a(u)},\;\;\;
\frac{\bar{b}^{\iota_1}{u}}{\bar{a}^{\iota_1}(u)}=
-\frac{\bar{b}(u)}{\bar{a}(u)}.
\end{equation} 

It is clear from the formulas (\ref{AB2}-\ref{AB4}) that the ratio
$\frac{b(u)}{a(u)}$ can be taken as a spectral parameter
$\frac{b(u)}{a(u)}=u$, $\frac{\bar{b}(u)}{\bar{a}(u)}=\bar{u}=
\theta-u.$

Finally, after appropriate normalization of $a(u)$ and $b(u)$ in
order to have $a(u)+b(u)=1$, one finds the following
solution of YBE
\begin{eqnarray}
\label{Sab}
a(u)&=&\frac{1}{1+u},\;\;\;\bar{a}(u)=\frac{1}{1+\theta-u},\;\;\;
a(\bar{u},v)=\frac{1}{1+\theta-u-v},\;\;\; \nonumber\\
b(u)&=&\frac{u}{1+u},\;\;\;\bar{b}(u)=\frac{\theta-u}{1+\theta-u}.\;\;\;
b(\bar{u},v)=\frac{\theta-u-v}{1+\theta-u-v}.
\end{eqnarray}

According to standard prescription of the $ABA$ technique the
logarithmic derivative of the transfer matrix at some point defines 
the Hamiltonian of the model
\begin{eqnarray}
\label{H1}
H=-\frac{\partial {\ln{\tau(u)}}}{\partial{u}}|_{u=0}.
\end{eqnarray} 

As it is known, in order Hamiltonian to be local, it is necessary
to have a value $u_0$, such that 
\begin{eqnarray}
\label{RI}
\check{R}_{i,j}(u_0)=I_{i,j}.
\end{eqnarray}
Analyzing the solutions of our $YBE$ (\ref{Sab}),
one can see from (\ref{R5}) that, at the point  $u_0=0$, only
$R_{i,j}(0)=I_{i,j}$ and $\bar{R}_{i,j}(0)=I_{i,j}$. 
As calculations shows the Hamiltonian is nevertheless local,
but it contains interaction between four neighbour points.

Technically, in order to calculate the logarithmic derivative
(\ref{H1}), one should put the expression of $\check{R}_{i,j}$
operators around $u_0=0$ up to linear terms
\begin{eqnarray}
\label{4RH}
R_{i,j}=I_{i,j}+uH_{i,j}, \nonumber\\
\bar{R}_{i,j}(u)=R_{i,j}(\theta)-uH_{i,j},\nonumber\\
R_{i,j}^{\iota_1}(u)=I_{i,j}-uH_{i,j},\nonumber\\
\bar{R}_{i,j}^{\iota_1}(u)={R}_{i,j}(-\theta)+uH_{i,j},
\end{eqnarray}
with
\begin{eqnarray}
\label{Hij}
H_{i,j}=\sum_{m,n}(-1)^{p(m)}{X_i}^{m}_{n}{X_j}^{n}_{m}
\end{eqnarray}
into the expression (\ref{M1}) of the Monodromy Matrix
\begin{eqnarray}
\label{MR} 
M(u)=\bar{R}_{01}^{\iota_1}(u)R_{12}(u)\bar{R}_{23}^{\iota_1}R_{34}(u)....
R_{12}^{\iota_1\iota_2}(u)\bar{R}_{23}^{\iota_2}(u)R_{34}^{\iota_1\iota_2}(u)\bar{R}_{45}^{\iota_2}
....
\end{eqnarray}

As a result, after some algebraic calculations, we will obtain the
following Hamiltonian for the present staggered t-J model
\begin{eqnarray}
\label{H2}
H&=&\theta \Delta \sum_{i=1}^{N}\left\{\sum_{\sigma=\uparrow \downarrow}
\left(
2-\frac{n_{i-1}}{2}-\frac{n_{i-2}}{2}\right)(c_{i,\sigma}^{+}c_{i+1,\sigma}
-c_{i+1,\sigma}^{+}c_{i,\sigma})\right.\nn\\
&+&\left.\sum_{\sigma=\uparrow\downarrow}
\left[\left(1-\frac{n_{i+1}}{2}+(-1)^{i}\theta\right)
c_{i+2,\sigma}^{+}c_{i,\sigma}
-\left(1-\frac{n_{i+1}}{2}-(-1)^{i}\theta\right)c_{i,\sigma}^{+}
c_{i+2,\sigma}\right]\right.\nn\\
&+&\left.2 \left[(\vec{S}_{i+2}+\vec{S}_{i-1})(\vec{S}_{i,i+1}-
\vec{S}_{i+1,i})+\vec{S}_{i+1}(\vec{S}_{i+2,i}-\vec{S}_{i,i+2})\right.\right.
\\
&+&\left.\left.\theta(-1)^{i}\left(\vec{S}_{i-1}\vec{S}_{i+1}-
\frac{1}{4}n_{i-1}n_{i+1}+
\frac{n_{i-1}+n_{i+1}}{2}\right)-i{\epsilon}^{abc}S_{i}^{a}S_{i+1}^{b}
S_{i+2}^c\right]\right\} \Delta,\nn
\end{eqnarray}
where ${\Delta}=\prod_{i=1}^{N}\Delta_{i}=\prod_{i=1}^{N}(1-n_{i\uparrow}
n_{i\downarrow})$ is the projector which excludes the double occupancy by
electrons at any site $i$.

The spin operators $\vec{S}_{i}$ and $\vec{S}_{i,j}$ are defined as
follows 
\begin{eqnarray}
\label{SS}
\vec{S}_{i}={\half}\Psi_{i,\beta}^{+}\vec{\sigma}_{\alpha}^{\beta}
\Psi_{i}^{\alpha}={\half}\Delta_{i}c_{i,\alpha}^{+}\vec{\sigma}
_{\beta}^{\alpha}c_{i}^{\beta}\Delta_{i}\nn\\
\vec{S}_{i,j}={\half}\Psi_{i,\beta}^{+}\vec{\sigma}_{\alpha}^{\beta}
\Psi_{j}^{\alpha}={\half}\Delta_{i}c_{i,\alpha}^{+}\vec{\sigma}
_{\beta}^{\alpha}c_{j}^{\beta}\Delta_{j},
\end{eqnarray}
where
\begin{eqnarray}
\label{PSI}   
\Psi^{1}=(1-n_{\uparrow})c_{\downarrow},
\Psi^{2}=c_{\uparrow}(1-n_{\downarrow}),
\end{eqnarray}
and $\vec{\sigma}$ are Pauli matrices.


\section{Algebraic Bethe Ansatz for the staggered t-J model.}
\indent

In this section we will apply the technique of $ABA$ \cite{FT,BBO,EK}
to the present model and find the eigenvalues and eigenstates of the 
Hamiltonian (\ref{H2}).

For this purpose let's introduce the $L$ operators as follows
\begin{eqnarray}
\label{L1}
(L_{i,j})_{k}^{k'}=\langle{k}\mid{R}_{i,j}\mid{k'}\rangle
\end{eqnarray}
which is a matrix in the horizontal auxiliary space and an operator 
in the quantum space. In matrix form it looks like
\begin{eqnarray}
\label{L2}
L_{i,j}=\left(
\begin{array}{lll}
a(u)-b(u)(1-n_{\uparrow})n_{\downarrow}& -b(u)c_{\uparrow}
c_{\downarrow}^{+}& b(u)(1-n_{\uparrow})c_{\downarrow}^{+}\\
 -b(u)c_{\uparrow}^{+}c_{\downarrow}& a(u)-b(u)n_{\uparrow}
(1-n_{\downarrow})& b(u)c_{\uparrow}^{+}(1-n_{\downarrow})\\
b(u)(1-n_{\uparrow})c_{\downarrow}& b(u)c_{\uparrow}
(1-n_{\downarrow})
&a(u)+b(u)(1-n_{\uparrow})(1-n_{\downarrow})
\end{array}
\right) \nonumber\\
\end{eqnarray}

The monodromy matrix $M_{k'}^{k}(u)$, which defined by matrix 
elements of the monodromy operators (\ref{M1},\ref{M2}) in the auxiliary 
space  can be expressed as a product  of $L_{i,j}$ matrices as follows
\begin{eqnarray}
\label{M3}
M_{0}(u)_{k'}^{k}=\langle{k}\mid{M}_{0}(u)\mid{k'}\rangle
=(-1)^{p(k)p(k')}(\bar{L}_{01}^{\iota_1})_{k_1}^k
(L_{02})_{k_2}^{k_1}....(L_{0N})_{k'}^{k_{N-1}}, \nonumber\\
M_{1}(u)_{k'}^{k}=\langle{k}\mid{M}_{1}(u)\mid{k'}\rangle
=(-1)^{p(k)p(k')}(L_{01}^{{\iota_1}{\iota_2}})_{k_1}^{k}
(\bar{L}_{02}^{\iota_2})_{k_2}^{k_1}....(\bar{L}_{0N}^{\iota_2})_{k'}^{k_{N-1}}.
\end{eqnarray}

Following the notations of the article \cite{EK}, one can write 
\begin{eqnarray}
\label{M4}
M_{s}(u)_{k'}^{k}=\left(
\begin{array}{lll}
A_{s,11}(u)\qquad &A_{s,12}(u)\qquad &B_{s,1}(u)\\
A_{s,12}(u)\qquad &A_{s,22}(u)\qquad &B_{s,2}(u)\\
C_{s,1}(u)\qquad &C_{s,2}(u)\qquad &D_{s}(u)
\end{array}
\right)\;\;\;\;
s=0,1
\end{eqnarray}
where $A_{s,ab}$, $B_{s,a}$, $C_{s,a}$, $D_{s}$; $(a,b=1,2)$ 
act on the quantum space.

Then, as a super-trace of the Monodromy matrix (\ref{M4}) the
transfer matrix (\ref{T1}) will have the form
\begin{eqnarray}
\label{T2}
\tau_{s}(u)=-A_{s,11}(u)-A_{s,22}(u)+D_{s}(u),\;\;\;\;s=0,1.
\end{eqnarray}  

This matrix elements of the Monodromy matrix are obeying the 
algebraic relations 
\begin{eqnarray}
\label{M5}
&&\hspace{-1cm}(-1)^{p(k'')(p(m')+p(m''))}
\check{R}_{k'm'}^{km}(u,v)M_{1,m''}^{m'}(u)M_{0,k''}^{k'}(v)\nn\\
&&\hspace{1cm}=(-1)^{p(k')(p(m)+p(m'))}
M_{1,m'}^{m}(v)M_{0,k'}^{k}(u)
\check{R}_{k''m''}^{k'm'}(u,v),\nonumber\\
&&\hspace{-1cm}(-1)^{p(k'')(p(m')+p(m''))}
\check{R}_{k'm'}^{km}(u,v)M_{0,m''}^{m'}(u)M_{1,k''}^{k'}(v)\nn\\
&&\hspace{1cm}=(-1)^{p(k')(p(m)+p(m'))}
M_{0,m'}^{m}(v)M_{1,k'}^{k}(u)\check{R}_{k''m''}^{k'm'}(u,v),
\end{eqnarray}
which are the consequence of $YBE$ (\ref{R3}-\ref{R4}).
In getting (\ref{M5}) we have used the properties (\ref{AB1}).

Consider now the empty fermionic state as a test ``vacuum''
\begin{eqnarray}
\label{O}
\mid{\Omega}\rangle_{s}=\mid{0,0,...,0}\rangle_{s}=\mid{0}\rangle_{1s}
\mid{0}\rangle_{2s}...\mid{0}\rangle_{Ns},\;\;\;s=0,1,
\end{eqnarray}
and let's check that $\mid{\Omega}\rangle$ indeed is a eigenstate
of transfer matrix (\ref{T2})
\begin{eqnarray}
\label{T3}
\tau_{s}(u)\mid{\Omega}\rangle_{s}=\nu_{s}^{(0)}\mid{\Omega}\rangle_{1-s}
\end{eqnarray}

{}From (\ref{M3}) ${\tau_{s}(u)}$ is a product
of $L_{i,j}$ matrices. Hence, in order to check (\ref{T3}) we should
first calculate $L_{0k}\mid{0}\rangle_{s,k}$. It appears that 
\begin{eqnarray}
\label{L3}
\bar{L}_{0k}^{\iota_2}\mid{0}\rangle_{s,k}=\left(
\begin{array}{lll}
\bar{b}(u)^{\iota_2}\qquad &0\qquad &\bar{b}^{\iota_2}(u)c_{k\downarrow}^{+}\\
0\qquad &\bar{b}^{\iota_2}(u)\qquad &\bar{b}^{\iota_2}(u)c_{k\uparrow}^{+}\\
0\qquad &0\qquad &\bar{a}^{\iota_2}(u)+\bar{b}^{\iota_2}(u)
\end{array}
\right)\mid{\Omega}\rangle_{k}\;\;\;\;
s=0,1.
\end{eqnarray}

We see that $L_{0k}\mid{0}\rangle_{k}$ is a upper-triangular 
matrix. Therefore the action of the product of $L_{0k}$ in the 
formula (\ref{M3}) on vacuum $\mid{\Omega}\rangle_{k}$ as a matrix 
will also have upper triangular form
\begin{eqnarray}
\label{M6}
M_{1}(u)_{k'}^{k}\mid{\Omega}\rangle_{1}&=&\left(
\begin{array}{lll}
[b^{{\iota_1}{\iota_2}}(u)\bar{b}^{\iota_2}(u)]^{\frac{N}{2}} &0 
&B_{1,1}(u)\\
0  & [b^{{\iota_1}{\iota_2}}(u)\bar{b}^{\iota_2}(u)]^{\frac{N}{2}} & B_{1,2}(u)\\
0 & 0 & 1
\end{array}
\right)\mid{\Omega}\rangle_{1},\;\;\;\nn\\
M_{0}(u)_{k'}^{k}\mid{\Omega}\rangle_{0}&=&\left(
\begin{array}{lll}
[\bar{b}^{\iota_1}(u){b}(u)]^{\frac{N}{2}} & 0 
&B_{0,1}(u)\\
0 & [\bar{b}^{\iota_1}(u){b}(u)]^{\frac{N}{2}} & B_{0,2}(u)\\
0 &0 & 1
\end{array}
\right)\mid{\Omega}\rangle_{0},\;\;\;
\end{eqnarray}
where we have used that $a(u)+b(u)=1$.

We see that the $B_{s,1}(u)$ and  $B_{s,2}(u), (s=0,1)$ 
operators create
one particle states while $C_{s,1}(u)$, $C_{s,2}(u)$ operators 
annihilate them
\begin{eqnarray}
\label{C1}
C_{s,a}(u)\mid{\Omega}\rangle_{s}=0,\;\;\;s=0,1;\;\;a=1,2
\end{eqnarray}
We see from the expression (\ref{M6}) that 
\begin{eqnarray}
\label{N1}
\nu_{s}^{(0)}(u)=1-2[b^{\iota_1}(u)\bar{b}(u)]^{\frac{N}{2}},
\end{eqnarray}
where $b^{\iota_1}(u)$ and $\bar{b}(u)$ defined by the equations (\ref{Sab})
and $N$ is the length of the chain.

This observation leads us to the following Ansatz for the eigenstates 
of $\tau(v)$ 
\begin{eqnarray}
\label{VF}
\mid{v_{1},v_{2},...v_{n}}\mid{F}\rangle_{0}=F^{{a_n}...{a_1}}
B_{0,a_1}(v_1)B_{1,a_2}(v_2)...
B_{0,a_n}(v_n)\mid{\Omega}\rangle_{0},\;\;\;a_i=1,2;\;\;
\end{eqnarray}
is a $n$ particle state. The $F^{{a_n}...{a_1}}$ is a function of 
spectral parameters $v_j$ to be specified later.

The action of the transfer 
matrix (\ref{T2}) on the states (\ref{VF}) is determined by the relations
(\ref{M6}) and the intertwining properties of the $A_{s,ab}(u)$,
$D_{s}(u)$, $B_{s,a}(u)$ operators defined from the $YBE$ (\ref{M5}).
The components of the intertwining relations, which we need 
for the construction of the $ABA$ are
\begin{eqnarray}
\label{DBA}  
D_{1}(u)B_{0,a}(v)&=&\frac{1}{b(u,v)}B_{1,a}(v)D_{0}(u)-
\frac{a(v,u)}{b(v,u)}B_{1,a}(u)D_{0}(v), \nn\\
A_{1,ba}(u)B_{0,c}(v)&=&\frac{r_{bc}^{b'c'}(u,v)}{b(u,v)}B_{1,c'}(v)A_{0,b'a}(u)
+
\frac{a(u,v)}{b(u,v)}B_{1,b}(u)A_{0,ca}(v),\nn\\
B_{1,a}(u)B_{0,b}(v)&=&r_{ab}^{b'a'}(u,v)B_{1,a'}B_{0,b'}(u),
\end{eqnarray}
where 
\begin{eqnarray}
\label{r}
r_{bc}^{b'c'}(v)=-a(v)\delta_{b}^{c'}\delta_{c}^{b'}+
b(v)\delta_{b}^{b'}\delta_{c}^{c'}=-a(v)I_{bc}^{c'b'}-b(v)\Pi_{bc}^{(1)b'c'}.
\end{eqnarray}
Here $\Pi^{(1),b'c'}_{bc}$ is a graded permutation operator for 
$p(1)=p(2)=1$, one can check that it fulfills the following $YBE$
\begin{eqnarray}
\label{HODVATS}
r(\lambda{-}\mu)_{a_{3}c_{3}}^{a_{2}c_{2}}r(\lambda)_{c_{2}d_{2}}^{{a_1}{b_1}}
r(\mu)_{a_{2}c_{2}}^{{d_2}{b_2}}=r(\mu)_{a_{2}c_{2}}^{a_{1}c_{1}}
r(\lambda)_{a_{3}b_{3}}^{c_{2}d_{2}}r(\lambda{-}\mu)_{d_{2}b_{2}}^{c_{1}b_{1}}.
\end{eqnarray}
Now by use of (\ref{DBA}), we can obtain, that the diagonal elements of the 
monodromy matrix act on the states (\ref{VF}) as follows
\begin{eqnarray}
\label{DVF}
D_{1}(u)\mid{v_{1},...,v_{n}}\mid{F}\rangle_{1} = \prod_{j=1}^{n}
\frac{1}{a(v_{j},u)}\mid{v_{1},...,v_{n}}\mid{F}\rangle_{0}+\nn\\
 +\sum_{k=1}^{n} {({\tilde{\Lambda}}_{k})}_{a_{1}...a_{n}}^{b_{1}...b_{n}}
B_{1,b_{k}}(u)\prod_{j=1,j\neq k}^{n}B_{b_{j}}
(v_{j})\mid{\Omega}\rangle_{0},\nn\\
\left[ A_{1,11}(u) + A_{1,22}(u)\right] \mid v_{1},...,v_{n} \mid{F}
\rangle_{1}= \nn\\
 = -\prod_{i=1}^{n}\frac{1}{b(u,v_{i})}[b^{t_{1}t_{2}}(u)\bar{b}^{t_{1}}(u)]
^{\frac{N}{2}}\tau_{a_{1}...a_{n}}^{(1)a'_{1}...a'_{n}}(u)
F^{a_{n}...a_{1}}\prod_{i=1}^{n}B_{a'_{i}}(v_{i})\mid{\Omega}\rangle_{0}+\nn\\
 + \sum_{k=1}^{n}(\Lambda_{k})_{a_{1}...a_{n}}^{b_{1}...b_{n}}
F^{a_{n}...a_{1}}
B_{1,b_{k}}(u)\prod_{i=1,j\neq k}^{n}B_{b_{j}}(v_{j})\mid{\Omega}\rangle_{0,}
\end{eqnarray}
where 
\begin{eqnarray}
\label{T4}
\tau_{a_{1}...a_{n}}^{(1)a'_{1}...a'_{n}}(u)&=&-
r_{ca_{1}}^{b_{1}a'_{1}}(u,v_{1})...r_{b_{n-1}a_{n}}^{ca'_{n}}(u,v_{n})\nn\\
&=&str[l_{n}(u,v_{n})l{(n-1)}(u,v_{n-1})...l_{1}(u,v_{1})],\;\;\;
\end{eqnarray}
and
\begin{eqnarray}
\label{l}
[l_{k}(u,v_{k})]_{b_{k-1}}^{b_{k}}=r_{b_{k-1}a_{k}}^{b_{k}a'_{k}}(u,v_{k}).
\end{eqnarray}

As it follows from (\ref{l}), $l_{k}$ is a $2\times 2$ matrix, the
elements of which are the operators 
\begin{eqnarray}
\label{l1}
l_{k}(u)=\left(
\begin{array}{ll}
l_{k,1}^{1}&l_{k,1}^{2}\\
l_{k,2}^{1}&l_{k,2}^{2}
\end{array}
\right)=\left(
\begin{array}{ll} 
b(u)I-a(u)e_{1}^{1}&-a(u)e_{1}^{2}\\
-a(u)e_{2}^{1}&b(u)I-a(u)e_{2}^{2}
\end{array}
\right).
\end{eqnarray}
where $e_{a}^{b}$ are quantum operators in the $n$-th space with matrix 
representation $(e_{a}^{b})_{\beta}^{\alpha}=\delta_{a}^{\alpha}
\delta_{\beta}^{b}.$

It is obvious, that the eigenvalue condition 
\begin{eqnarray}
\label{nu}
(D_{s,1}(u)-A_{s,11}(u)-A_{s,22}(u))\mid v_{1},...v_{n}\mid F\rangle_{s}=
\nu_{s}(u,v_{1},...,v_{n})\mid v_{1},...,v_{n}\mid F\rangle_{1-s}\;\;\;
\end{eqnarray}
will be fulfilled if 

$i)$ we impose the cancellation of unwanted terms in (\ref{DVF})
\begin{eqnarray}
\label{i}
[(\tilde{\Lambda}_{k})_{a_{1}...a_{n}}^{b_{1}...b_{n}}-
({\Lambda}_{k})_{a_{1}...a_{n}}^{b_{1}...b_{n}}]F^{a_{n}...a_{1}}=0
\end{eqnarray}
called Bethe equations $(BE)$, and

$ii)$ we solve the eigenvalue problem for the small transfer matrix (\ref{T4})
\begin{eqnarray}
\label{ii}
\tau_{a_{1}...a_{n}}^{(1)a'_{1}...a'_{n}}(u;v_{1},...,v_{n})F^{a_{1}...a_{n}}
=\nu^{(1)}(u,v_{i})F^{a'_{1}...a'_{n}},
\end{eqnarray}
then we have the following expression for eigenvalues 
\begin{eqnarray}
\label{nu2}
\nu_{1}(u;v_{1},...,v_{n})&=&\prod_{i=1}^{n}\frac{1}{b(v_{i},u)}+
[b^{t_{1}t_{2}}(u)\bar{b}^{\iota_2}(u)]^{\frac{N}{2}} \prod_{j=1}^{n}
\frac{1}{b(u,v_{j})}\nu^{(1)}(u,v_{i}),\nn\\ 
\nu_0(u;v_{1},...,v_{n})&=&\prod_{i=1}^{n}\frac{1}{b(v_{i},\bar{u})}+
[b^{t_{2}}(\bar{u})\bar{b}^{t_{2}}(\bar{u})]^{\frac{N}{2}}\prod_{j=1}^{n}
\frac{1}{b(\bar{u},v_{j})}\nu^{(1)}(\bar{u},v_{i}).
\end{eqnarray}

For the solution of second equation $ii)$ we should make the $ABA$ for a 
small auxiliary problem of chain, with length $n$ (number of particles) 
and ``nested'' transfer matrix $\tau_{a_{1}...a_{n}}^{(1)a'_{1}...a'_{n}}
(u;v_{1},...,v_{n}).$ This is why all this procedure is called Nested 
Algebraic Bethe Ansatz (NABA). In the article \cite{EK}, it was demonstrated
how to calculate $\Lambda_{a_{1}...a_{n}}^{b_{1}...b_{n}}$ and
$\tilde \Lambda_{a_{1}...a_{n}}^{b_{1}...b_{n}}$ and to reduce the condition 
of cancellation of the  unwanted terms for the ordinary $t-J$ model 
to some equation. It is not necessary to repeat the same calculation 
here since it differs very little from the carried one. The only 
difference is appearing in the term 
\begin{eqnarray}
\label{AA}
(A_{1,11}+A_{1,22})\mid \Omega\rangle=[b^{t_{1}t_{2}}(u)\bar{b}^{t_{2}}(u)]
^{\frac{N}{2}}\mid \Omega\rangle,
\end{eqnarray}
therefore we obtain the following conditions
\begin{eqnarray}
\label{TF}
\tau_{b_{1}...b_{n}}^{(1)b'_{1}...b'_{n}}(v_{k}\mid v_{1},...,v_{n})
F^{b_{n}...b_{1}}=[b^{t_{1}t_{2}}(v_{k})\bar{b}^{\iota_2}(v_{k})]^{-\frac{N}{2}}
\prod_{i=1,i\neq{k}}^{n}\frac{b(v_{k}v_{i})}{b(v_{i}v_{k})}
F^{b'_{n}...b'_{1}},
\end{eqnarray}
as a Bethe equations.

In the next step of the $NABA$ we should find the  eigenvalues and 
eigenstates of $\tau^{(1)}(u)$. It is clear from the equations 
(\ref{HODVATS}) that we have another small integrable model with the 
$R$ matrix $r_{ab}^{a'b'}(u)$ defined by the formula (\ref{r}) and 
the corresponding transfer matrix $\tau^{(1)}(u).$

Therefore we should apply a non ordinary $ABA$ to this problem.
The $YBE$ for the problem is  
\begin{eqnarray}
\label{rMM}
r_{ab}^{a'b'}(u,v)\hat{M}_{a'}^{(1)a''}(u)\hat{M}_{b'}^{(1)b''}(v)=
\hat{M}_{b}^{(1)b'}(v)\hat{M}_{a}^{(1)a'}(u)r_{a'b'}^{a''b''}(u,v)
\end{eqnarray}
where $M_{a}^{(1)a'}$ is the corresponding (nested) Monodromy
matrix.

Now if we define
\begin{eqnarray}
\label{M7}
M^{(1)}(u)=\left(
\begin{array}{ll}
A^{(1)}(u)&B^{(1)}(u)\\
C^{(1)}(u)&D^{(1)}(u)
\end{array}
\right),\;\;\tau^{(1)}(u)=-A^{(1)}(u)-D^{(1)}(u),
\end{eqnarray}
then by use of the formula (\ref{r}) and $YBE$ (\ref{rMM}), we find
\begin{eqnarray}
\label{DBA1}
D^{(1)}(u)B^{(1)} (v)&=&\frac{1}{b(u,v)}B^{(1)} (v)D^{(1)}(u)+
\frac{a(v,u)}{b(v,u)}B^{(1)} (u)D^{(1)}(v),\nn\\
A^{(1)}(u) B^{(1)}(v)&=&\frac{a(u,v)}{b(u,v)}B^{(1)}(u) A^{(1)}(v)+
\frac{1}{b(v,u)}B^{(1)}(v) A^{(1)}(u),\nn\\
B^{(1)}(u) B^{(1)}(v)&=&B^{(1)}(v) B^{(1)}(u).
\end{eqnarray}

Let's take as reference state 
\begin{eqnarray}
\label{O2}
\mid 0\rangle_{k}^{(1)} =\left(
\begin{array}{l}
1\\
0
\end{array}
\right),\;\;\nn\\
\mid\Omega\rangle^{(1)}=\mid 0\rangle_1 ^{(1)}
\dots\mid 0\rangle_n ^{(1)}=\bigotimes_{k=1}^n\mid 0\rangle_k ^{(1)}.
\end{eqnarray}

The action of the nested monodromy matrix $M^{(1)} (u)$ on the 
reference state $\mid\Omega\rangle^{(1)}$ is described by the action $l_k (u)$ 
on $\mid 0\rangle_k ^{(1)}$, which we can find from (\ref{l1}). So we
obtain
\begin{eqnarray}
\label{AD2}
A^{(1)}(u)\mid\Omega\rangle^{(1)}=\prod_{i=1}^n [b(u,v_i )-a(u,v_i )]
\mid\Omega\rangle^{(1)}=\prod_{i=1}^{n}\frac{b(u,v_i )}{b(v_i ,u)}\mid\Omega
\rangle^{(1)},\nn\\
D^{(1)}(u)\mid\Omega\rangle^{(1)}=\prod_{i=1}^n b(u,v_j )\mid\Omega
\rangle^{(1)}.
\end{eqnarray}

For the eigenstates of $\tau^{(1)}(v)$, we are going to do the following 
Ansatz
\begin{eqnarray}
\label{VF2}
\mid v_{1}^{(1)},\dots,v_m^{(1)}\rangle=B^{(1)}(v_1 ^{(1)})B^{(1)}(v_2 ^{(1)})
\dots B^{(1)}(v_m ^{(1)})\mid\Omega\rangle^{(1)}.
\end{eqnarray}
The action of $\tau^{(1)}(u)$ on the states (\ref{VF2}) is the same 
as the action of the diagonal elements of (\ref{M7}) on that states.
By use of (\ref{DBA}) we will obtain
\begin{eqnarray}
\label{AD3}
D^{(1)}(u)\mid v_1^{(1)},\dots ,v_m^{(1)}\rangle 
&=&\prod_{j=1}^n b(u,v_j)\mid 
v_1^{(1)},\dots ,v_m^{(1)}\rangle \nn\\
&+&\sum_{k=1}^m\Lambda_k^{(1)}B^{(1)}(u)
\prod_{i=1,i\neq k}^mB^{(1)}(v_i)\mid\Omega\rangle^{(1)},\nn\\
A^{(1)}(u)\mid v_1^{(1)},\dots ,v_m^{(1)}\rangle
&=& \prod_{i=1}^{m}\frac{1}{b(v_i^{(1)},u)}
\prod_{j=1}^n\frac{b(u,v_j)}{b(v_j,u)}
\mid v_1^{(1)},\dots ,v_m^{(1)}\rangle\nn\\
&+&\sum_{k=1}^m\tilde{\Lambda}_k^{(1)}B^{(1)}(u)\prod_{j=1,j\neq k}
B^{(1)}(v_j) \mid\Omega\rangle^{(1)}.
\end{eqnarray}

{}From the expression (\ref{AD3}) we can easily write the eigenvalues 
of $\tau^{(1)}(u)$
\begin{eqnarray}
\label{tt}
\tau^{(1)}\mid v_1^{(1)},\dots ,v_m^{(1)}\rangle=&-&\left[
\prod_{i=1}^m\frac{1}{b(v_i^{(1)},u)}
\prod_{j=1}^n\frac{b(u,v_j)}{b(v_j,u)}+\right.\nn\\
&+&\left.\prod_{i=1}^m\frac{1}{b(u,v_j^{(1)})}\prod_{j=1}^nb(u,v_j)\right]\mid 
v_1^{(1)},\dots ,v_m^{(1)}\rangle.
\end{eqnarray}

One can get simply the first set of Bethe equations by comparing (\ref{tt})
with the formula (\ref{TF}). Inputing $u=v_k$ in (\ref{TF}) we obtain
\begin{eqnarray}
\label{bb1}
[b^{\iota_1}(v_k)\bar{b}(v_k)]^{\frac{N}{2}}=\prod_{i=1}^mb(v_i^{(1)},v_k),\;\;
\;k=1,2,\dots ,n.
\end{eqnarray}

The second set of Bethe equations, which are the conditions of cancellation of
the unwanted terms $\Lambda_k^{(1)}$ and $\tilde{\Lambda}_k^{(1)}$ are 
similar to the corresponding equations of the standard $XXX$ model and can 
be found easily as 
\begin{eqnarray}
\label{bb2}
\prod_{j=1}^nb(v_j,v_k^{(1)})=\prod_{i\neq k}\frac{b(v_k^{(1)},v_i^{(1)})}
{b(v_i^{(1)},v_k^{(1)})},\;\;\;k=1,2,\dots,m.
\end{eqnarray}
This is exactly the same equation as found in \cite{EK}.

Finally we find
\begin{eqnarray}
\label{nu3}
\nu_1(u,\{v_i\})&=&\prod_{i=1}^n\frac{1}{b(v_i,u)}-
[b^{\iota_1}(u)\bar{b}(u)]^\frac{N}{2}
\prod_{j=1}^n\frac{1}{b(u,v_j)}\left[\prod_{i=1}^m\frac{1}{b(v_i^{(1)},u)}
\prod_{j=1}^n\frac{b(u,v_j)}{b(v_j,u)}\right.\nn\\
&+&\left.\prod_{i=1}^m\frac{1}{b(u,v_i^{(1)})}
\prod_{j=1}^nb(u,v_j)\right],\nn\\
\nu_0(u,\{v_i\})&=&\prod_{i=1}^n\frac{1}{b(v_i,\bar{u})}-
[\bar{b}^{\iota_1}(u) b(u)]^\frac{N}{2}
\prod_{j=1}^n\frac{1}{b(\bar{u},v_j)}
\left[\prod_{i=1}^m\frac{1}{b(v_i^{(1)},\bar{u})}
\prod_{j=1}^n\frac{b(\bar{u},v_j)}{b(v_j,\bar{u})}\right.\nn\\
&+&\left.\prod_{i=1}^m\frac{1}{b(\bar{u},v_i^{(1)})}
\prod_{j=1}^nb(\bar{u},v_j)\right] 
\end{eqnarray}
as the $n$ particle state eigenvalues of transfer matrices $\tau_1(u)$ 
and $\tau_0(u)$ respectively.

But the transfer matrix of our staggered model is the product of
$\tau_0(u)$ 
and $\tau_1(u)$, therefore the eigenvalues $\nu(u,\{v_i\})$ 
of $\tau(u)$ are
\begin{eqnarray}
\label{nu4}
\nu(u,\{v_i\})=\nu_0(u,\{v_i\})\nu_1(u,\{v_i\})
\end{eqnarray}
with the Bethe equations (\ref{bb1}) and (\ref{bb2}) unchanged.
 
Let us now to calculate the energy of excitations over the test ``vacuum''
$\mid\Omega\rangle$, called bare energy, which will be dressed
in a real ground state due to interactions with particles in a filled
Dirac sea. The bare energy is a logarithmic derivative of eigenvalues
(\ref{nu3}) and (\ref{nu4}) at the point $u=0$. The simple calculation
gives the energy and the momentum of $n$-particle state 
$\mid{v_1,...v_n}\mid{F}\rangle$ as it follows
\begin{eqnarray}
\label{EE}
E_0(\{v_j\})&=&-\sum_{j=1}^{n}\left\{\frac{1}{v_j^2 +1/4}-
\frac{1}{(v_j-\theta)^2 +1/4}\right\},\nn\\
iP(\{v_j\})&=&\sum_j^n\left\{\log\frac{v_j+1/2}{v_j -1/2}+
\log\frac{v_j-\theta+1/2}{v_j-\theta -1/2}\right\},
\end{eqnarray}
where we have redefined the spectral parameters as $v_j \rightarrow
v_j-1/2$.

The solution of the $BE$ (\ref{bb1}) and (\ref{bb2}) is usually obtained in 
the thermodynamic limit ($N,n,m \rightarrow \infty$, with the fixed ratio
${n \over N}, {m \over N})$. In this case instead of a discrete set of
spectral parameters $v_j$ one introduces the distribution of continuous 
density $\rho(v)$ of rapidities. The ground state is defined by filling
up the Dirac sea(s) of negative energies by the electrons. It was 
argued in the article \cite{BBO} that the ground state of the $t-J$ model
is defined by the string solutions of length two, which are filling of 
all states with negative energy. The lowest energy value can be reached
by maximal filling of negative energy states, which corresponds to
${n \over N}=1$ and with zero magnetization, corresponding to
$m={n \over 2}$. In our model it is clear from the expression
of the energy (\ref{EE}) that only the spectral parameters of the
interval
\begin{equation}
\label{uu}
-\infty < u < {\theta \over 2}
\end{equation}
have to be filled in order to form a ground state. But since this is
exactly equal to half of lattice sites $N$, we will have a ground 
state corresponding to quarter filling of all states.

After introducing the densities, the $BE$ becomes an integral
equation of the form \cite{BBO}
\begin{eqnarray}
\label{TBA1}
\pi \rho(v)+ \int_Q^{\infty}du \frac{\rho(u)}{(v-u)^2+1}=
{1 \over 2}\left\{\frac{1}{v^2 +1/4}+
\frac{1}{(v-\theta)^2 +1/4}\right\} \;, 
\end{eqnarray}
where $Q$ defines the rapidity of the Fermi level. In case of
quarter filling it is equal to zero.

The energy of ground state is defined by the equation
\begin{eqnarray}
\label{TBA2}
E_0&=&-2N \int_Q^{\infty}\left\{\frac{1}{v^2 +1}-
\frac{1}{(v-\theta)^2 +1}\right\}\nn\\
&=&-2N (\rho(0)-\rho(\theta)).
\end{eqnarray}
At usual half filling $Q=-\infty$, and as it follows from the
expression (\ref{TBA2}), $E=0$.

It seams to us, that this model provides an interesting possibility
to analyze by means of exact integrability the physics of systems
with the quarter filled ground state.

\section{Acknowledgments}
A.S. would like to acknowledge INTAS grant 0524 and T.S. INTAS grant
99-1459 for partial financial support.

\end{document}